\documentstyle[prl,aps,floats,psfig]{revtex}
\tighten
\begin{document}
\large


\title{ Black Holes with Zero Mass}


\author{Ulises Nucamendi$^{\,a,b,}$\footnote{e-mail: ung@star.cpes.susx.ac.uk} \\
\large {and} \\
\large {Daniel Sudarsky$^{\,b,}$\footnote{e-mail: sudarsky@nuclecu.unam.mx}} \\
\vspace*{0.4cm}
\small {\it $^{a}$Centre for Theoretical Physics}\\
\small {\it University of Sussex}\\
\small {\it Brighton BN1 9QJ, Great Britain.}\\
\vspace*{0.4cm}
\small {\it $^{b}$Instituto de Ciencias Nucleares} \\ 
\small {\it Universidad Nacional Aut\'onoma de M\'exico} \\ 
\small {\it Apdo. Postal 70-543 M\'exico 04510 D.F, M\'exico}. }
\maketitle
\vspace*{0.4cm}


\centerline {\bf abstract}

We consider the spacetimes corresponding to static
Global Monopoles with interior boundaries corresponding to
a Black Hole Horizon and analyze the behavior of the appropriate
ADM mass as a function of the horizon radius $r_H$.
We find that for small enough $r_H$, this mass
is negative as in the case of the regular global monopoles,
but that for large enough $r_H$ the mass becomes positive
encountering an intermediate value for which we have a Black Hole with
zero ADM mass.

\vskip 1.5cm

\section{Introduction}
\label{sec:Introduction}

Global monopoles are topological defects that arise in certain theories
where a global  symmetry is spontaneously broken.
 The simplest and most studied example is the $O(3)$ model, in which
one finds that
the
 static, spherically symmetric solutions, have in general an energy
density that decreases for large distances as $1/r^2$\cite{Vilenkin}.
This would lead in a Newtonian analysis to a divergent expression for
the total mass. When we turn to the general relativistic analysis,
this problem translates in the fact that the resulting spacetime is
not asymptotically flat, and thus the standard ADM mass is not well defined.
The effect of the $1/r^2$  behavior of the  density on the spacetime is that
the latter develops a deficit angle at large distances The fact that,
at small distances, the behavior deviates from that,  results in the
appearance
of a small phenomenological "core mass" which turns out to be negative in
all cases considered \cite{Harari}.
Moreover, an analysis of the behavior of geodesics in the large distance
regime does indeed support such interpretation of this core mass because
its effect turns out to be repulsive\cite{Harari}.
This ``core mass" is then evidently not the standard ADM mass.
The question of what exactly  one is talking about when referring to this
core mass has been resolved in \cite{Ulises} through the application
of the standard type of Hamiltonian analysis to the class of spacetimes that
are Asymptotically-flat-but-for-a-deficit-angle (AFDA$\alpha$) \cite{Ulises}.

For these (AFDA$\alpha$) spacetimes one can also define future and past
conformal
null infinity, and thus the notion of a black hole and of its horizon.
In fact solutions corresponding to global monopoles with such interior
horizons have been found in \cite{Liebling}, \cite{Maison}.

In this paper we study the behavior of the ADM mass of these AFDA$\alpha$
spacetimes
as a function of the horizon area, concentrating in particular on its sign,
which we find changes in the regime where one would interpret as going
from a situation that would be naturally described as a ``black hole
inside a monopole core" to that which would be naturally described as
a ``black hole with a global monopole inside".

We shall adhere to the following conventions
on index notation in this paper: Greek indices
($\alpha$, $\beta$, $\mu$, $\nu$,...) range
from $0$ to $3$, and denote tensors on
(four-dimensional) spacetime. Latin indices,
alphabetically located after the letter i
(i,j,k,...) range from $1$ to $3$, and denote tensors on a
spatial hipersurface $\Sigma$; whereas
Latin indices,
from the beginning of the alphabet (a,b,c,d,...)
range from $1$ to $3$,
and denote  indices in the
internal space of the scalar fields.
 The  metric  for the internal space is just  the flat Euclidean metric $
\delta_{ab}$.
The signature of the spacetime
metric {\bf g} is
$ (-,+,+,+)$. Geometrized units, for which
$G_N=c=1$ are used in this paper.

\medskip
\section{The Global Monopole Spacetime}
\medskip

The theory of a scalar field with spontaneously broken internal $O(3)$
symmetry, minimally coupled to gravitation, is described by the action:
\\
\begin{equation}
S = \int \sqrt{(-g)}
[(1/16\pi) R - (1/2)(\nabla^{\mu}\phi_a)
(\nabla_{\mu} \phi^a) -V(\phi)]  d^{4} x.
\label{(1.1)}
\end{equation}
\\
where R is the scalar curvature of the
spacetime metric, $\phi_a$ is a
triplet of scalar fields, and
$V(\phi)$, is potential depending only
on the magnitude $\phi =( \phi_a \phi^a)^{1/2}$,
which we will take to be the ``Mexican Hat"
$V(\phi) = (\lambda/4)
(\phi^2 - \eta^2)^2$.  \\
\\
The gravitational field equations following from
the Lagrangian (\ref{(1.1)}) can be written as
\\
\begin{equation}
\label{rmunu}
R^{\mu\nu} - {1\over 2} g^{\mu\nu}R = 8\pi T^{\mu\nu}
\end{equation}
\\
where
\\
\begin{eqnarray}
\label{Teff}
T^{\mu\nu}_{\rm sf} &=& \nabla^\mu\phi^a\nabla^\nu\phi_a
- g^{\mu\nu}\left[{1\over 2}  (\nabla \phi^a)(\nabla \phi_a) +
V(\phi)\right] \ .
\label{tsf}
\end{eqnarray}
\\
The equation of motion for the scalar fields become
\\
\begin{eqnarray}
\Box \phi^a = {\partial V(\phi)\over \partial \phi_a} \ .
\label{seq}
\end{eqnarray}
\\
We are interested in spacetimes with topology
$\Sigma \times R$, where
$\Sigma $ has the topology of $(R^3 - B) \cup C$,
with $B $ a 3-ball, and $C$ a compact manifold with
$S^2$ boundary.

We will focus on the sector corresponding to the
asymptotic behavior characteristic of the
Hedgehog ansatz:

\begin{equation}
\phi^a
\approx  \eta  x^a /r.
\label{(1..2)}
\end{equation}
\\
where the $x^a$  are  asymptotic Cartesian
coordinates.
Within this sector, there is a static,
spherically symmetric solution \cite{Vilenkin}
with metric
given by:
\\
\begin{equation}
 ds^2 = - B(r) dt^2 + S(r) dr^2
 + r^2  ( d\theta^2
+ \sin^2 (\theta) d\varphi^2 ),
\label{solution}
\end{equation}
\\
and scalar field
\begin{equation}
\phi^a = \eta f(r)  x^a /r
\label{escalar}
\end{equation}
\\
and with the following asymptotic behavior of,
\\
\begin{equation}
B \approx S^{-1} \approx 1 - \alpha - 2 M/r + O(1/r^2),
\quad  f \approx 1 + O(1/r^2)
\label{CA}
\end{equation}
\\
where $\alpha= 8\pi \eta^2$.
Redefining the $r$ and $t$ coordinates
as $r \to (1-\alpha)^{1/2} r$ and
$t \to (1-\alpha)^{-1/2} t$, respectively,
and defining $\widetilde M = M (1-\alpha)^{-3/2}$,
we obtain the asymptotic form for the metric:
\\
\begin{equation}
 ds^2 = - (1-2\widetilde M /r) dt^2 +
(1-2\widetilde M /r)^{-1} dr^2
 +(1-\alpha) r^2  ( d\theta^2
+ \sin^2 (\theta) d\varphi^2 ).
\end{equation}
\\
As we mentioned it is natural to associate the
parameter $\widetilde M$  with the mass of the
configuration because it can be seen that the proper
acceleration of the  ($\theta,\varphi, r$) = constant
world lines is $a= - \widetilde M/(r(r-2\widetilde M))$.
However, as we explained before it is not the standard ADM mass.
This is also evident from the fact that in the specific solution
$\widetilde M$ turns out to be negative \cite{Harari}, while the matter
certainly satisfy the dominant energy condition,
under which the ADM mass of a regular solution would be positive
\cite{yau}, \cite{witten}.

These issues are clarified by the introduction of concept of
asymptotically-flat-but-for-a-deficit-angle spacetimes
(A.F.D.A $\alpha$) and the
standard asymptotically-flat-but-for-a-deficit-angle spacetime
(S.A.F.D.A $\alpha$).
The ADM mass of any spatial hipersurface of the former being defined in
terms of its spatial metric and the metric of a particular slice of the
S.A.F.D.A $\alpha$ spacetime (see \cite{Ulises} for details).
\\
In fact we take as the S.A.F.D.A $\alpha$ spacetime the metric,
\\
\begin{equation}
ds^2 = - dt^2 + dr^2 + (1-\alpha) r^2 (d\theta^2
+ \sin^2\theta d\varphi^2) \,\,\,\,\,.
\label{RGMSconst1}
\end{equation}
\\
The natural generalization of the ADM mass for the
A.F.D.A $\alpha$ spacetimes is
\\
\begin{equation}
16 \pi (1-\alpha) M_{ADM\alpha} =
\int_{\partial\Sigma}
dS_{i} ( h^{0ij} h^{0kl} - h^{0ik} h^{0jl} )
 D_{k}^0 (h_{jl}).
\label{masa}
\end{equation}
\\
where $h_{ij}$ is the spatial metric of the slice
of the spacetime for which the ADM$\alpha$ mass will be evaluated,
$h_{ij}^{0}$ is the spatial metric of a static slice of the
S.A.F.D.A $\alpha$ spacetime,
and $D_{j}^{0}$ the covariant derivative associated with the latter
(in fact, it looks just like the usual ADM formula, but with the
quantities associated with the flat metric replaced by the
S.A.F.D.A.$\alpha$ metric), and just like this,
it is the numerical value of the true Hamiltonian
( a true generator of ``time translations");
so, it is natural to interpret this as the mass
(or energy) of the A.F.D.A.$\alpha$ spacetimes.
\\
Let us write a static spherically symmetric metric in the form
\\
\begin{equation}
ds^2 = - \left(1-\frac{2m}{r}\right) e^{-2\delta} dt^2 +
\left(1-\frac{2m}{r}\right)^{-1} dr^2 + (1-\alpha) r^2 (d\theta^2
+ \sin^2\theta d\varphi^2) \,\,\,\,\,.
\label{RGMS}
\end{equation}
\\
where the functions $m(r)$ and $\delta(r)$ depend only on the radial
coordinate $r$.
\\
We also introduce the following dimensionless quantities
\begin{eqnarray}
\tilde r &:=& r\cdot \eta\lambda^{1/2}  \,\,\,\,,\\
\tilde m &:=& m\cdot \eta\lambda^{1/2} \,\,\,\,,\\
\label{rotilde}
\tilde E &:=& E\cdot \frac{1}{\eta^{2}\lambda}  \,\,\,\,,\\
\alpha &:=& 8\pi \eta^2  \,\,\,\,,
\end{eqnarray}
\\
Introducing the metric (\ref{RGMS}) and the ansatz (\ref{escalar})
in the expressions (\ref{rmunu}) and (\ref{seq}), we obtain the
final form of the  equations of motion :
\\
\begin{eqnarray}
\label{massu}
\partial_{\tilde r} \tilde m &=&  4\pi \tilde r^2 \tilde E -
\frac{\alpha}{2(1 - \alpha)}\,\,\,\,\,,\\
\label{lapsefi}
\partial_{\tilde r} \delta &=& - \frac{\alpha\tilde r}{2}
(\partial_{\tilde r}f)^2,  \,\,\,\, \\
\label{scalarfi}
\partial_{\tilde r\tilde r} f &=&
-\left[ \frac{2}{\tilde r} - \partial_{\tilde r} \delta -
2 A^2
\left(4\pi \tilde r \tilde E - \frac{\alpha}{2(1 - \alpha)\tilde r}
-\frac{\tilde m}{\tilde r^2}\right)
\right] (\partial_{\tilde r}f) \nonumber \\
& & + A^2
\left[f(f^2-1) + \frac{2f}{(1 - \alpha)\tilde r^2} \right], \,\,\,\,\,
\end{eqnarray}
where
\begin{eqnarray}
A^2 = \left( 1 - \frac{2\tilde m}{\tilde r} \right)^{-1},
\end{eqnarray}
and
\begin{eqnarray}
\tilde E &=&
\frac{\alpha}{8\pi}
\left[\frac{(\partial_{\tilde r} f)^2}{2A^2} +
\frac{f^2}{(1 - \alpha)\tilde r^2} + \frac{(f^2-1)^2}{4} \right].
\end{eqnarray}
\\
In this paper, we will make use of the fact that when the metric
is written as (\ref{RGMS}), the formula (\ref{masa}) for the ADM$\alpha$
mass of the spacetime can be expressed as:
\\
\begin{equation}
M_{\rm ADM\alpha} = M = {\rm lim}_{r\rightarrow\infty} \,\,
m( r)   \,\,\,\,.
\end{equation}
\\
Regular solutions require the following boundary conditions at the origin
$f(0)=0$, $\tilde m(0)=0$, $\tilde m(0)_{,\tilde r}=- \alpha / (2(1 - \alpha))$
 and using
a standard shooting method to compute $f(0)_{,\tilde r}$. The solutions
are found by the standard one parameter
shooting method \cite{Num}. This is possible because the equation for $\delta$
decouples, and can thus be solved
independently,  after the rest of the functions are solved, and
because, once the function $f$
approaches $1$ (sufficiently fast) at $\infty$, eq.(\ref{massu}) ensures that
$\tilde m $ converges to a finite
value.  \\
\\
A very useful tool in the analysis of this kind of configurations
is provided by the consideration of the limit in which $\lambda \to \infty$
which can formally be taken by
replacing the potential term by the constraint $\phi^2 = \eta^2$.
\\
In fact in this case the field equations become;
\\
\begin{equation}
R^{\mu\nu} - {1\over 2} g^{\mu\nu}R = 8\pi T^{\mu\nu}_{\rm const}
\label{rmunucons}
\end{equation}
\\
where
\\
\begin{eqnarray}
\label{Teffcons}
T^{\mu\nu}_{\rm const} &=& \nabla^\mu\phi^a\nabla^\nu\phi_a
- g^{\mu\nu}\left[{1\over 2}  (\nabla \phi^a)(\nabla \phi_a)  \right] \ .
\label{tsfcons}
\end{eqnarray}
\\
while the equation of motion for the scalar fields become
\\
\begin{eqnarray}
\Box \phi^a = \frac{1}{\eta^2} \phi^b (\Box \phi_b) \phi^a  \ .
\label{seqcons}
\end{eqnarray}
\\
Using the metric (\ref{RGMS}) we can verify that the  anzats
$\phi_a = \eta x_a /r$, satisfies
identically  eq. (\ref{seqcons})  and the eqs. (\ref{rmunucons}) reduce to
\\
\begin{eqnarray}
\label{massuconst}
\partial_{\tilde r} \tilde m &=& 0    \,\,\,\,\,,\\
\label{lapseficonst}
\partial_{\tilde r} \delta &=& 0  \,\,\,\,,
\end{eqnarray}
and
\begin{eqnarray}
\tilde E &=&
\frac{1}{8\pi}
\left( \frac{\alpha}{(1 - \alpha)\tilde r^2}  \right).
\end{eqnarray}
\\
Which results in the solution given by the metric
\\
\begin{equation}
ds^2 = - \left(1-\frac{2M}{r}\right) dt^2 +
\left(1-\frac{2M}{r}\right)^{-1} dr^2 + (1-\alpha)(r^2d\theta^2
+ r^2\sin^2\theta d\varphi^2) \,\,\,\,\,.
\label{RGMSconst}
\end{equation}
\\
which for the choice $M=0$, is in fact what is taken as the
S.A.F.D.A $\alpha$ spacetime (\ref{RGMSconst1}).

\medskip
\section{Solutions with Black Hole Horizons}
\medskip

We note that  eq. (\ref{RGMSconst}) with $M>0$ corresponds to  the
 case of a solution with a regular event horizon.
In this case, as in the standard asymptotically flat case,
the mass is in fact positive and given by
$M = \left(\frac{A_H}{16\pi (1-\alpha)}\right)^{1/2}$
where $A_H$ is the area of the horizon.

The issue is then, whether
in the general situation, i.e. without taking the limit
$\lambda \to \infty$, we will find also positive masses, or whether the
presence of the monopole will in some instances dominate and make the mass
negative?. Intuition of course suggests that the latter will be the case,
and that the two regimes (positive and negative mass) are
 possible, with the interplay between the two scales, the monopole core
scale given by $r_c \equiv (\eta\lambda^{1/2})^{-1}$,
and the black hole radius given by $r_H $, defining which
one prevails. In the case $r_H > r_c$, which we will consider
as a ``monopole within a black hole`", one expects the black hole features
to dominate, and that therefore the $M_{\rm ADM\alpha}$ will be positive.
In the case $r_H < r_c$, which we will consider as a ``black hole inside
a monopole", one expects the monopole features to dominate,
and thus that the  $M_{\rm ADM\alpha}$ will be negative.
Furthermore there should exist
a specific regime where $r_H \approx r_c $ and for which the two
tendencies will exactly compensate each other so that the mass
should be zero.

We will investigate these issues numerically and will find that the above
picture is in fact confirmed by the results.

\begin{figure*}
\vspace{1cm}
\psfig{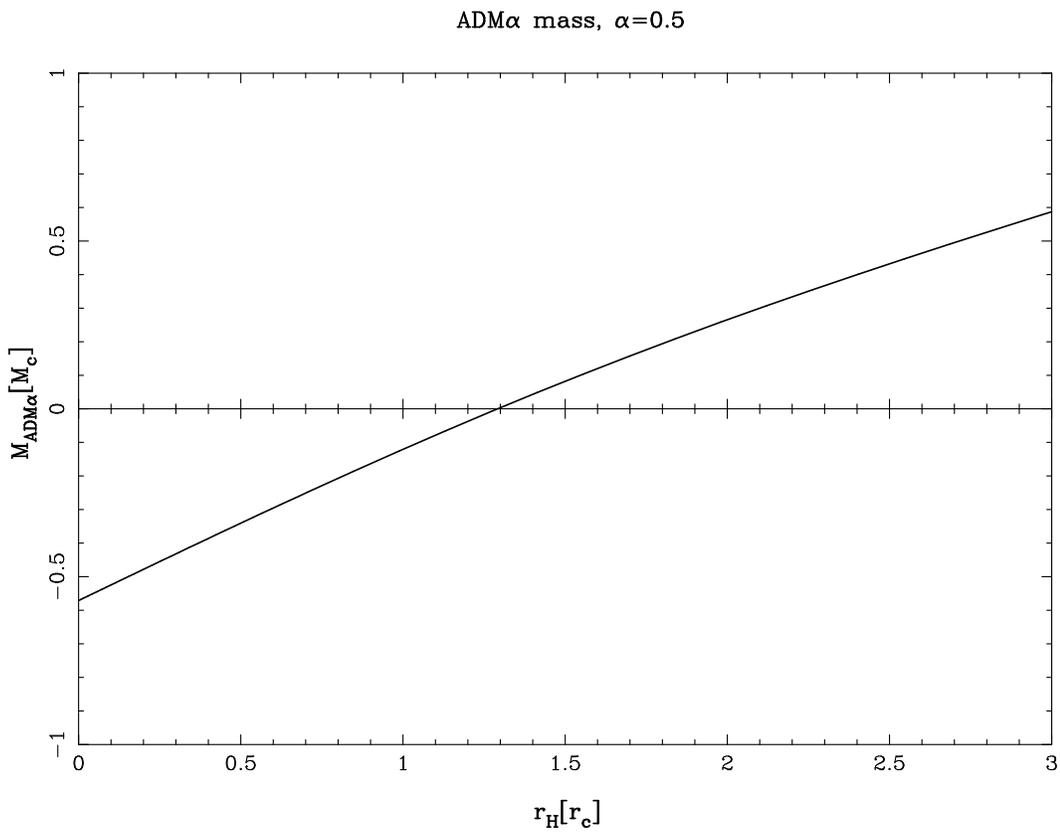}
\hspace*{-2.5in}
\caption[]{\label{f:masa}
The ADM$\alpha$ mass vs the horizon radius. Here $\alpha = 0.5$
and $M_c = r_c \equiv (\eta\lambda^{1/2})^{-1}$.}

\end{figure*}

\begin{figure*}
\vspace{1cm}
\psfig{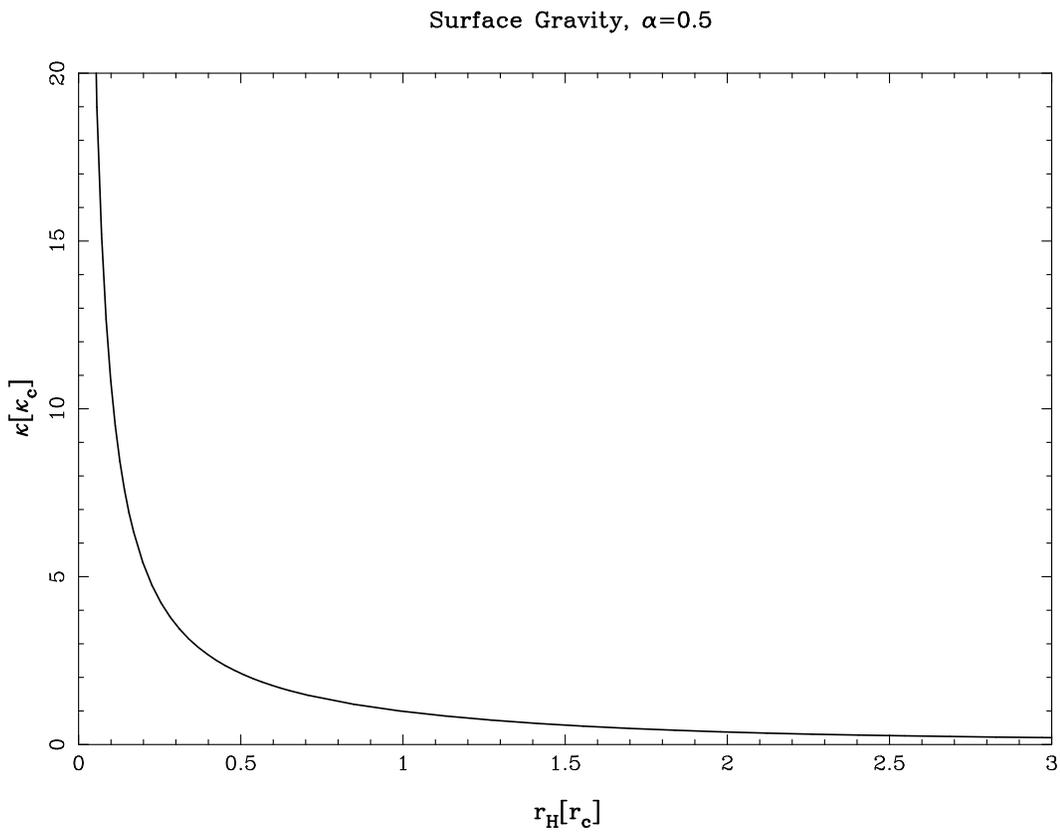}
\hspace*{-2.5in}
\caption[]{\label{f:grasup}
The surface gravity vs the horizon radius. Here 
$\alpha = 0.5$ and 
$\kappa_c = r_c^{-1} \equiv (\eta\lambda^{1/2})$.}
\end{figure*}

Configurations corresponding to static spherically symmetric regular
black hole horizons with area $A= 4\pi r_H^2= {{ 4 \pi \tilde
r_H^2}\over{\lambda \eta^2}}$ are those that satisfy
the standard boundary conditions at infinity, i.e.,
${\rm lim}_{r\rightarrow\infty} f(r)=1$,
${\rm lim}_{r\rightarrow\infty} m(r)=M$, where $M$ is a constant,
 and that at
$\tilde r=\tilde r_H$ satisfy:
\begin{eqnarray}
\label{cfm}
2\tilde m(\tilde r_H)= \tilde r_H \,\,.
\end{eqnarray}
We ensure that the time-translational Killing field of the metric is
normalized to unit at infinity. This is done by fixing the constant that
appears in the integration of the equation
for $\delta$ in such a way that ${\rm lim}_{r\rightarrow\infty}
\delta(r)=0$. \\
\\
The value of the
monopole field at
$\tilde r=\tilde r_H$, $f(\tilde r_H)=f_{H}$,
is taken as the shooting  parameter,
and is thus fixed by the boundary conditions at spatial infinity.
The derivatives at $r=r_H$ of the functions $f(r)$ and $m(r)$ are
given by:
\\
\begin{eqnarray}
\tilde m(\tilde r_H)_{,\tilde r}= \frac{\alpha(f_H^2-1)}{2(1-\alpha)}
\left[ 1 + \frac{(1-\alpha) \tilde r_H^2 (f_H^2 - 1)}{4} \right] \,\,,
\label{derimasa}
\end{eqnarray}
\\
\\
\begin{eqnarray}
f(\tilde r_H)_{,\tilde r}=
\frac{f_H (2 + \tilde r_H^2 (1 - \alpha)(f_H^2-1))}
{\tilde r_H(1-\alpha)(1-2m(\tilde r_H)_{,\tilde r})}
\,\,,
\end{eqnarray}
\\
In the same way as in the regular case the system is analyzed as a
standard one dimensional shooting problem.\\
\\
In the $\lambda \to \infty$ limit we can see from (\ref{RGMSconst}) that
the ADM mass is always positive, and  this can be easily
understood if we note that this limit can be thought
to represent the extreme case of a monopole inside a black hole . \\
\\
The Fig.\ref{f:masa} shows the behavior of the ADM$\alpha$ mass as a
function of the horizon radius. Note that as expected there exists a
radius where the mass ADM$\alpha$ vanishes.
These are therefore zero mass black holes.  \\
\\
We must emphasize that in this class of spacetimes the mass is not
 positive definite and therefore there is nothing really paradoxical
 about the fact that there are black holes that have zero mass. However we
 must also point out that the mass in this case is not just a definition as
it really reflects the effects on the test particles at large distances
from the "body" which in the cases treated here are the monopole core and/or
the black hole horizon. In this case a zero value for the mass means that
at large distances, the proper acceleration of the
static bodies (i.e., those following integral curves of the static Killing
field) falls off faster than $1/r^2$. Moreover, as pointed out
in \cite{Ulises}
these black holes satisfy the standard laws of black hole dynamics
and are nondegenerate as can be seen from the evaluation of the
surface gravity $\kappa$. This is obtained from the expression
$t^{\mu} \nabla_{\mu} t^{\nu} = \kappa t^{\nu}$,
i.e., the surface gravity
is defined in terms of the  acceleration at the horizon of the
time-translational Killing field $t^{\mu}$ which is unit
at infinity. In general for a spherically symmetric system the
surface gravity is \cite{visser},
\\
\begin{equation}
\kappa =(\eta \lambda^{1/2}) \frac{1}{2\tilde r_{\rm H}}e^{-\delta(\tilde
r_{\rm H})}\left[
1-2 \tilde m(\tilde r_H)_{,\tilde r} \right]\label{ksss}
\label{surfaceGravity}
\end{equation}
\\
here the derivative of the metric function evaluated at the
horizon can be evaluated from formula (\ref{derimasa}).
The surface gravity is positive definite as it is shown
in Figure \ref{f:grasup}.  \\
\\
Previous works on this system (e.g.\cite{Maison}) were unable
to study the main point of this work because they lacked an
appropriate definition of the mass for the class of A.F.D.A.$\alpha$
spacetimes. \\
\\
These type  of models with unusual asymptotic are a very interesting
ground to investigate the robustness of the standard results of the
physics of black holes and to understand which of those results are
specific to the asymptotic flatness assumption. This point might seem
to be of purely academic interest, but we must remember that our universe
is not asymptotically flat, and that the latter is just an
approximation that is introduced in order to simplify the
treatment of regions of
spacetime that might be regarded as ``isolated" from the rest of the
universe, thus the issue of the degree to which the standard results
are independent of the precise form of the asymptotics is indeed of
practical importance because it  will tell us which of them might
have to be taken as really pertinent
to our universe. In this work we have learned for example that the mass of
a black hole need not be positive and might indeed be negative or even
zero.  Here it is worth recalling that, as a result of the fact that
the spacetimes in question {\it are not
asymptotically flat}, the notion of ADM mass used throughout this work is
different from the usual one, and thus,  in particular,  the standard theorems
concerning the positivity  of the standard ADM mass do not apply  (for
details see\cite{Ulises}).
Other results that we feel should be studied in this context
include, for example, the black hole
uniqueness theorems, the black hole entropy results arising from the
various
proposals for a quantum theory of gravity, etc. Some of these issues are
currently under investigation and will be the subject of forthcoming
articles.

\vskip 1cm

\vskip 1cm

\break
\section*{Acknowledgments}

 This work was in part supported by
DGAPA-UNAM grant No IN121298 and  by CONACyT
grant 32272-E. U.N. is supported by a CONACyT Postdoctoral 
Fellowship Grant 990490.

\end{document}